\documentclass[preprint,12pt]{elsarticle}

\usepackage{amssymb}

\usepackage{color}

\newcommand{\Z}{\mathbb{Z}}
\newcommand{\beq}{\begin{equation}}
\newcommand{\eeq}{\end{equation}}
\newcommand{\R}{\mathbb{R}}

\newtheorem{definition}{Definition}

\journal{Signal Processing}

\begin{document}

\begin{frontmatter}



\title{Causal conditioning and instantaneous coupling in causality graphs}


\author[gipsa,melb]{Pierre-Olivier Amblard} 
\ead{bidou.amblard@gipsa-lab.inpg.fr}
\author[gipsa]{Olivier J.J. Michel}
\ead{olivier.michel@gipsa-lab.inpg.fr}


\address[gipsa]{GIPSAlab/CNRS UMR 5216/ BP46, \\ 38402 Saint Martin d'H\`eres cedex, Grenoble,  France}
\address[melb]{ The University of Melbourne, Dept. of Math\&Stat., Parkville, VIC, 3010, Australia}

\begin{abstract}
The paper    investigates   the link between Granger causality graphs recently formalized by Eichler and   directed information theory developed by Massey and Kramer. We particularly insist on the implication of two  notions of  causality that may   occur    in physical   systems.   It is well accepted that   dynamical causality   is    assessed by the conditional transfer entropy, a measure appearing naturally as a part of directed information.  Surprisingly   the notion of instantaneous causality is often overlooked, even if it was clearly understood in early works. In the bivariate case, 
 instantaneous   coupling    is measured adequately by the instantaneous information exchange, a measure that   supplements   the transfer entropy in the decomposition of directed information. 
   In this paper, the focus is put on the multivariate case and conditional graph modeling issues. In this framework,    we show that the decomposition of directed information into the sum of transfer entropy and information exchange does not hold anymore. Nevertheless, the discussion allows to put forward the two measures as pillars for the inference of causality graphs. We illustrate this on two synthetic examples   which allow us to discuss not only the theoretical concepts, but also    the practical estimation issues.  

\end{abstract}


\end{frontmatter}

\sloppy


\section{Introduction}

\subsection{Motivations}
Graphical modeling has received a major attention in many different domains such as neurosciences \cite{JordS01}, econometry \cite{Eich07}, complex networks \cite{Newm10}. It proposes a representation paradigm for explaining how information flows between the nodes of a graph. The graph vertices are in most cases, and in particular in this paper, associated to  synchronous time series. Inferring a graph thus requires to define edges or links between the vertices. Granger \cite{Gran80,Gran88} proposed a set of axiomatic definitions for the causality between say $x$ and $y$ (with a slight abuse of notation, each vertex will be named after its associated time series). Granger's definitions are  based on the improvement that observations of $x$ up to some time $t-1$ may provide for predicting $y$ at $t$.  The fundamental idea in Granger's approach is that {\em past and present may cause the future but the future cannot cause the past} \cite[Axiom A]{Gran80}. Granger's work also stresses the importance of side information, accounting for the presence of all other vertices but $x$ and $y$, for assessing the existence of a link between two nodes. This leads to what will be referred to as bivariate case (absence of side information) or multivariate case (presence of side information) in the sequel. The use of Granger causality in the latter case is mainly due to Eichler and Dahlaus \cite{DahlE03,Eich07,Eich11}.

In \cite{Eich11}, precise definitions of Granger causality graphs are presented, and the two notions of dynamical causality and instantaneous causality (as we call them in this paper) are put forward. Note that the notion of instantaneous causality was present in the early works on Granger causality, but this notion, which seems quite weak compared to the other, has been overlooked in the modern studies on causality, especially in the applications.
Instantaneous dependence in complex networks may arise from different origins. Actually, if one cannot easily conceive instantaneous information exchange between nodes, the recording process (including filters, sample and hold devices, converters) contains integrators over short time lags. Any information flowing between two nodes within a delay shorter than the integration time may then be seen as instantaneous. Such a case is often met in systems requiring long integration times per sample, as for example in fMRI. Alternately, instantaneous coupling may occur if  noise contributions in structural models are no longer independent. 


\subsection{Aim of the paper and outline}
The purpose of this paper is to provide a new insight in the problems related to instantaneous coupling, and to show how the presence of such coupling may affect the estimated structure of a graphical model that should provide a sparse representation of a complex system. The paper is focused on the interplay of two types of causality: dynamical causality and instantaneous causality, overlooked in all other works on directed information.  The possibility to estimate directed information based measures with $k$-nearest neighbors based tools is illustrated.  

We  begin the paper by a short review of possible approaches to Granger causality. 
Section III   then introduces  a brief review and some definitions of Eichler and Dalhaus  causality graphs \cite{DahlE03,Eich11} and presents an enlightening toy problem, where instantaneous coupling strongly affects the edge detection in a graphical model. Theoretical relations exhibiting the link between directed information theory and Granger causality graphs  are developed in the following section. The last section discusses some practical implementation issues and gives a full treatment of the toy problem studied previously.

\section{Approaches to Granger causality} 
\subsection{Model based approaches}
In Geweke's pioneering work \cite{Gewe82,Gewe84} an autoregressive modeling approach (for the bivariate as well as for the multivariate case) was adopted in order to provide a practical implementation of Granger causality graphs. Such a model based approach motivated further studies : Information theoretic tools were also added by Rissanen and Wax \cite{RissW87}, in order to account for the regression model complexity. Directed transfer functions, that are frequency domain filter models for  Granger causality, were derived in \cite{Gran80,Gran88} for neuroscience applications.  Nonlinear extensions have been  proposed \cite{GourBF06}, with recent developments relying on functional estimation in RKHS \cite{MariPS08,AmblMRH12}. All these approaches  are intrinsically parametric, and as such, may introduce some bias in the analysis. 

\subsection{Information theoretic based measures}
An alternative for assessing the existence of a link between nodes was early elaborated in the bivariate case (see for example a sample from the literature \cite{SaitH81,GourMR87,Schr00,PaluV07,Solo08}). It consists in adapting  information theoretic measures such as mutual information or information divergences to assess the existence and/or strength of a link between two nodes. The motivations for introducing such tools rely upon the ability of information theoretic measures to account for the entire probability density function of the observations (provided that such  a density exists), instead of only second order characteristics as for linear filter modeling approaches. Among these references, one of the oldest and may be the less known was developed by Gouri\'eroux {\it et.al. } \cite{GourMR87}   where a   generalization of Geweke's idea \cite{Gewe82} using Kulback divergences is introduced. It is  noteworthy  that the tools they introduced was   later rediscovered  by Massey and Kramer in their development of bivariate directed information theory. 

The development of directionality or causality specific measures was initiated by Marko's work on directed information \cite{Mark73}, and extended by Massey \cite{Mass90}, and later Kramer \cite{Kram98} who introduced causal conditioning by side information. This offers a means to account for side information, or to tackle the multivariate case.  First steps in exploring the relation between Geweke's approach of Granger causality and directed information theoretic tools were made in \cite{AmblM09} for the Gaussian case and further insights are developed in  \cite{AmblM11:it}, or in \cite{QuinKC11} in the absence of instantaneous dependence structure. In  \cite{AmblM11:it}, a directed information based new definition is  proposed for Granger causality. Eichler's recent paper \cite{Eich11} studies this latter issue in a graph modeling framework either from a theoretical point of view recoursing to probability based definitions, or in a parametric modeling context.

\section{Causality graphs}

We briefly review the notion of causality graph as developed by Eichler. The main reference is  \cite{Eich11}
where a complete presentation of causality graphs as well as a study of their Markovian properties are developed.

\subsection{Definitions}

Let $x_V=\{x_V(k),  k\in \Z\}$ be a $d$-dimensional discrete time stationary multivariate process on some probability space. The probability measures 
are assumed to be absolutely continuous with respect to Lebesgue measure, and  their  density  
associated to it  will be noted $P$.
$V$ is the index set $\{1,\ldots,d\}$. For $a\in V$ we  denote  $x_a$ as the corresponding component of $x_V$. Likewise,  for any subset $A\subset V$, $x_A$ is the corresponding multivariate process. The information obtained by observing $x_A$ up to time $k$ is resumed by the filtration generated by
$\{ x_A(l), \forall l\leq k\}$. It is denoted as $x_A^k$.

 Following \cite{Gran80,Gran88,Eich11}, three definitions may be proposed for Granger causality. The first one is   based on simple forward prediction,    the root concept underlying Granger causality.
The two  next definitions correspond to alternative choices in defining 
 instantaneous causality  .
Let $A$ and $B$ be two disjoint subsets of $V$. Let $C=V\backslash (A\cup B)$. 

\begin{definition}[Dynamical] \label{GrangerDyn:def}
 $x_A$ does not (dynamically) cause $x_B$ if for all $k\in \Z$,
\begin{eqnarray*}
P\big( x_B(k+1)  \big| x_A^k,x_B^k,x_C^k \big)  = P\big( x_B(k+1)  \big| x_B^k,x_C^k  \big)
\end{eqnarray*}
\end{definition}
\vspace{.4cm}

\noindent Dynamical Granger causality states that $x$ causes $y$ if the prediction of $y$ from its past is improved when also considering the past of $x$. Moreover, this is relative to any side information observed {\it prior} to the prediction.  This is the meaning of definition \ref{GrangerDyn:def}: Conditional to its past and to the side information, $x_B$ is independent of the past of $x_A$. In mathematical terms, $x_B^k \longrightarrow x_A^k \longrightarrow x_B(k+1)$ is a Markov chain  conditionally to the side information ($x_C^k$). 

Conditioning on   $x_C^{k}$ instead of 
 $x_C^{k+1}$ in def. \ref{GrangerDyn:def} raises an important issue: In a model estimation framework not aimed at identifying links between possibly  all pairs of nodes, one may think about accounting for the present of $x_C$ in the prediction problem; this is for instance the case for ARMA modeling. However, conditioning on $x_C^{k+1}$ weakens the effectiveness of the definition of causality by introducing a symmetry in the causal relationship between $B$ and $C$. Conditioning is therefore restricted to the past of the observation, in a strict sense.  
 This excludes the possibility of instantaneous dependences, for which  a separate definition is required.  There are however two possible definitions.

\begin{definition}[Instantaneous] \label{GrangerInst:def}
$x_A$ does not (instantaneously) cause $x_B$ if for all $k\in \Z$,
\begin{eqnarray*}
P\big( x_B(k+1)  \big|  x_A^{k+1},x_B^k,x_C^{k+1} \big)  = P\big( x_B(k+1)  \big| x_A^{k},x_B^k,x_C^{k+1} \big)
\end{eqnarray*}
 \end{definition}
 \vspace{.4cm}

\noindent
The second possibility is the following.
\begin{definition}[Unconditional instantaneous] \label{GrangerInstUn:def}
$x_A$ does not (unconditionally instantaneously) cause $x_B$ if for all $k\in \Z$,
\begin{eqnarray*}
P\big( x_B(k+1)  \big|  x_A^{k+1},x_B^k,x_C^{k} \big)  = P\big( x_B(k+1)  \big| x_A^{k},x_B^k,x_C^{k} \big)
\end{eqnarray*}
 \end{definition}
 \vspace{.4cm}

\noindent
Firstly,   definitions \ref{GrangerInst:def} and \ref{GrangerInstUn:def} are easily shown to be symmetrical in $A$ and $B$ (application of Bayes theorem). Secondly, taking as side information $x_C^{k+1}$ in def. \ref{GrangerInst:def}  instead of $x_C^{k}$ in def.  \ref{GrangerInstUn:def} 
is fundamental here. If the side information is considered up to time $k$ only, the instantaneous dependence or independence is not conditional to the remaining nodes in $C$. In fact inclusion of all the information up to time $k$ in the conditioning variables allows to  instantaneously test dependence or independence between $x_A(k+1)$ and $x_B(k+1)$. The independence  tested is not conditional if $x_C(k+1)$ is not included in the conditioning set, whereas the independence tested is conditional if  $x_C(k+1)$ is included. Thus  the choice is crucial when dealing with the type of  graph of instantaneous dependence obtained. In definition  \ref{GrangerInst:def} the  graphs obtained are conditional dependence graph as usual in graphical modeling \cite{Whit89,Laur96}. On the contrary, graphs obtained with definition   \ref{GrangerInstUn:def} are dependence graph which do not have the  nice Markov properties that conditional dependence graphs may have. 

The two possible types of causality (dynamical or instantaneous) will be encoded on the graphs by two different types of edges between vertices. Dynamical causality will be represented by an arrow, hence symbolizing directivity, whereas instantaneous causality will be represented by a line.

\subsection{A Detailed example}
\label{detailedexample:ssec}

\begin{figure}[p]
\begin{center}
 \includegraphics[scale=.6]{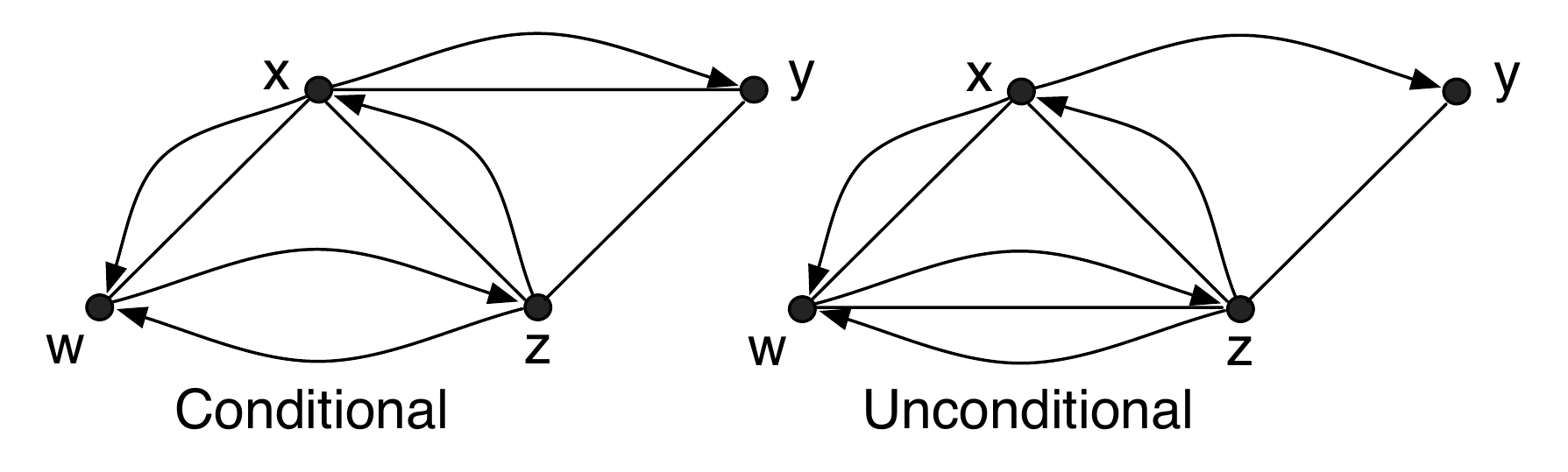}
 \end{center}
 \caption{Causality graphs for the example developed in the text. Illustration of the difference between the two definitions of instantaneous causality. }
 \label{jouet:fig}
  \end{figure}

For the sake of illustration we consider a four dimensional simple example. Let $\rho_{1,2,3} \in (-1 ,1) $ and let 
\begin{eqnarray}
\label{covariance:eq}
\Gamma_\varepsilon =
 \left(
 \begin{array}{cccc}
 1 & \rho_1 & 0& \rho_1\rho_2\\
   \rho_1  & 1 & 0  &\rho_2 \\
   0 & 0 & 1 & \rho_3\\
  \rho_1\rho_2&    \rho_2 &   \rho_3  &   1  
 \end{array}
 \right)
\end{eqnarray}
 be the covariance matrix of the  i.i.d. zero mean Gaussian sequence $(\varepsilon_{w,t} ,\varepsilon_{x,t} ,\varepsilon_{y,t} ,\varepsilon_{z,t} )^\top$. The inverse of $\Gamma_\varepsilon$, known as the precision matrix, reveals the conditional independence relationship between the components of the noise (since it is Gaussian), and reads
\begin{eqnarray}
\label{precision:eq}
\Gamma_\varepsilon^{-1} =
 \left(
 \begin{array}{cccc}
 d_1 & -d_1\rho_1 & 0&0 \\
  -d_1\rho_1  & d_1d_2(1-\rho_1^2\rho_2^2-\rho_3^2) &d_2 \rho_2\rho_3  &-d_2 \rho_2 \\
   0 &  d_2 \rho_2\rho_3   & d_2 (1-\rho_2^2) & -d_2 \rho_3\\
0 &  -d_2 \rho_2 &   -d_2 \rho_3  &   d_2  
 \end{array}
 \right)
 \end{eqnarray}
where $d_1=1/(1-\rho_1^2),d_2=1/(1-\rho_2^2-\rho_3^2)$.
Consider the following structural model
\begin{eqnarray*}
\left\{
 \begin{array}{ccl}
 w_t &=& f_w(w_{t-1},x_{t-1},z_{t-1}) + \varepsilon_{w,t} \\
 x_t &=& f_x(x_{t-1},z_{t-1})+ \varepsilon_{x,t} \\
 y_t &=& f_y(x_{t-1},y_{t-1}) + \varepsilon_{y,t} \\
 z_t &=& f_z(w_{t-1}, z_{t-1}) + \varepsilon_{z,t} \\
 \end{array}
 \right. 
 \label{exemple:eq}
\end{eqnarray*}
To infer the causality graph, we first look for directed link  between pairs of nodes. In such a structural model, if a signal $\alpha$ at time $t$ depends through the function $f_\alpha$ on another signal $\beta$ at time $t-1$, then there is a link $\beta \longrightarrow \alpha$. For example, consider the question of whether there is a link from $z$ to $w$ or not? We have from the definition of the model 
\begin{eqnarray*}
P(w_t | w^{t-1},z^{t-1}, (x,y)^{t-1})&=& P_{\varepsilon_w}\big( w_t - f_w(w_{t-1},x_{t-1},z_{t-1})\big) \\
P(w_t | w^{t-1}, (x,y)^{t-1})&=& E_{z^{t-1}} \left[ P_{\varepsilon_w}\big( w_t - f_w(w_{t-1},x_{t-1},z_{t-1})\big) \right]
\end{eqnarray*}
which are obviously not equal here. Therefore $z \longrightarrow w \big| € x,y$. Consider now the case of $z$ and $y$. We have 
$P(y_t | y^{t-1},z^{t-1}, (x,w)^{t-1})= P_{\varepsilon_y}\big( y_t - f_y(x_{t-1},y_{t-1})\big)=P(y_t | y^{t-1}, (x,w)^{t-1})$. Thus 
$z \not\longrightarrow y \big| € x,w$. Doing this pairwise or following the intuitive point of view described above leads to the set of oriented edge depicted in the causality graph in figure (\ref{jouet:fig}). To get the instantaneous edges, as discussed in the previous section, we have two possible definitions. If side information is considered up to time $t-1$, we obtain the unconditional graph in figure (\ref{jouet:fig}). Indeed for the unconditional graph, testing for the presence of an edge between $x$ and $y$, we evaluate
$P(x_t \big| x^{t-1},y^t, (w,z)^{t-1})= P(\varepsilon_x \big| \varepsilon_y)= P(\varepsilon_x)$ since $\varepsilon_x$ and $\varepsilon_y$ are independent (examine $\Gamma_\varepsilon$ and remember the noises are Gaussian). Note that doing this for all pairs, we really obtain the graph of  dependence relationships. For the conditional graph, we instead evaluate
$P(x_t \big| x^{t-1},y^t, (w,z)^{t})= P(\varepsilon_x \big| \varepsilon_y, \varepsilon_w, \varepsilon_z)$. In this case, we really measure the conditional dependence between $x$ and $y$. It turns out in the example that even if independent, $\varepsilon_x$ and $\varepsilon_y$ are  dependent conditionally to $\varepsilon_z$, and therefore there is an undirected edge between $x$ and $y$ in the conditional graph. 

\section{Directed information and causality graphs}

We start with a brief reminder on the main definitions of directed information and some related results. Bivariate analysis results are sketched, to provide better insight in discussing the multivariate case. 

Massey's work focusses on information measures for system that may exhibit  feedback  \cite{Mass90} . In this framework, Massey proved that the appropriate information measure was no longer the mutual information but the directed information.
For two subsets $A$ and $B$, directed information is defined  by
\begin{eqnarray*}
I( x_A^k \rightarrow x_B^k )  = \sum_{i=1}^k I( x_A^i ; x_B(i) | x_B^{i-1} )
\end{eqnarray*}
where $ I( x_A^i ; x_B^{i} | x_B^i )$ stands for the usual conditional mutual information \cite{CoveT93}.  
Later in \cite{Kram98}, Kramers   introduced  the idea of causal conditioning and defined the causally conditioned entropy as a modified version of the Bayes chain rule for conditional entropy: While the usual chain rule  writes 
$H(x_B^k | x_A^k) = \sum_{i=1}^k H\left(x_B(i) \big|x_B^{i-1},  x_A^k   \right)$, causally conditioned entropy is defined as
\begin{eqnarray*}
 H(x_B^k  \| x_A^k) = \sum_{i=1}^k H\left( x_B(i) \big|x_B^{i-1},  x_A^i  \right).
\end{eqnarray*}
The difference lies on the conditioning on $x_A$  which is now considered up to time $i$ only for each term entering the sum. 
From the definitions above, the directed information is easily decomposed into the difference of two terms 
 $$I( x_A^k \rightarrow x_B^k )   = H( x_B^k)- H(x_B^k  \| x_A^k)$$
 which could be compared to the well known (sometimes admitted as a definition) formula for the mutual information  
 $I( x_A^k ; x_B^k )   = H( x_B^k)- H(x_B^k  | x_A^k)$. 
 
 Assuming the presence of side information, causal conditioning of directed information is thus defined by substituting causally conditioned entropies to entropies in the definition of directed information. Causally conditioned directed information is given by
\begin{eqnarray}
I( x_A^k \rightarrow x_B^k  \big\|x_C^k  ) \!\!\!&=&\!\!\! H( x_B^k\big\| x_C^k) - H( x_B^k \big\| x_A^k , x_C^k) \nonumber \\
 \!\!\!&=&\!\!\! \sum_{i=1}^k I\left( x_A^i ; x_B(i) | x_B^{i-1} ,x_C^i  \right). \label{infodircausal:eq}
\end{eqnarray}

From these definitions, Massey and Kramer derived two interesting results.  The first one is the following equality, where $Dx_A^k = (0, x_A^{k-1}) $ represents the delayed (one time lag) version of $x_A$ : 
\begin{eqnarray}
I(x_A^k \rightarrow x_B^k ) + I(x_B^k \rightarrow x_A^k )  &=  & I( x_A^k ; x_B^k )   
+ I(x_A^k \rightarrow x_B^k \big\| Dx_A^k) \label{sum-dir-info}
\end{eqnarray}
This implies that the sum of the directed information is larger than the mutual information. The term  $ I(x_A^k \rightarrow x_B^k \big\|Dx_A^k)$ is positive (sum of positive contributions) and accounts for the  instantaneous information exchange.  Using equation 
(\ref{infodircausal:eq}) one easily gets 
\begin{eqnarray}
 I(x_A^k \rightarrow x_B^k \big\|€ Dx_A^k)  = \sum_{i=1}^k I\left( x_A(i) ;x_B(i) \big| x_B^{i-1} , x_A^{i-1}  \right) \label{eii:eq}
\end{eqnarray}
which is symmetric with respect to $A$ and $B$. 

It is noteworthy that by its definition, directed information accounts for instantaneous information exchange as well as for dynamical information exchange. Then, in the sum of the directed information in the l.h.s. of equation (\ref{sum-dir-info}), the contribution of instantaneous information is counted twice. It is   counted only once in the mutual information, and this explain the remaining term in the r.h.s. of the equation. \\

The instantaneous information exchange term $I(x_A^k \rightarrow x_B^k \big\| Dx_A^k)$ is zero if and only if $I\left( x_A(i) ;x_B(i) \big| x_B^{i-1} , x_A^{i-1}  \right) =0, \forall i$, {\it i.e.} $x_A$ and $x_B$ are independent conditionally on their past. Such a situation may occur for multivariate Markov processes described by $ X_V(t)=f(X_V(t-1))+ \epsilon_V(t)$, where $\epsilon_V(t)$ is an i.i.d. multivariate noise process with independent components. Note that in the example of the preceding section, $\Gamma_\varepsilon$ is not diagonal, therefore the noise components are correlated and lead  to some instantaneous information exchanges between some nodes (in a non trivial way).  


We are at this point ready to examine how directed information may be used in causality graphs. In front of multivariate measurements, two approaches are possible. The first one is a bivariate analysis in which we study directed information between pairs of nodes, forgetting the side information (remaining nodes). The second one accounts for side information but will need some more developments. Even if the bivariate framework is a naive approach, it is   presented since it gives some insights on how directed information is applied. We then turn to the more tricky multivariate analysis.


\subsection{Bivariate analysis in graphs}
Consider two disjoint subsets $A$ and $B$ of $V$.
 The directed information may be re-expressed as the sum
\begin{eqnarray}
I(x_A^k \rightarrow x_B^k)\!\! &=& \!\!I(x_A^k \rightarrow x_B^k || €  Dx_A^k)+ I(Dx_A^k \rightarrow x_B^k ), \nonumber
\end{eqnarray}
where the first term is the instantaneous information exchange defined by equation (\ref{eii:eq}), whereas the second 
$I(Dx_A^k \rightarrow x_B^k )$ will be referred to as  the {\em transfer entropy}, following  Schreiber definition proposed in a different framework  \cite{Schr00}. In the absence of any side information, these terms account for  the instantaneous causality and for the  dynamical causality respectively. Indeed, the transfer entropy reduces to zero if and only if $I(x_A^{i-1} ; x_B(i) | € x_B^{i-1})=0$,   $\forall i$ or equivalently if and only if $x_A$ does not dynamically cause $x_B$ (see def. \ref{GrangerDyn:def}). 
Furthermore, we have seen above that $I(x_A^k \rightarrow x_B^k || €  Dx_A^k)=0$ if and only if $x_A$ and $x_B$ are independent conditionally on their past, or in the words of our definitions, if and only if  $x_A$ does not instantaneously cause $x_B$. This result extends those obtained in the Gaussian bivariate case in \cite{AmblM09} and in \cite{BarnBS09} restricted to the dynamical causality. Again, these conclusions hold in the  sole case where no side information is considered.

\subsection{Multivariate analysis in graphs}

It is assumed in the sequel than the set of measurements or nodes $V$ is   partitioned   into three disjoint subsets $A$, $B$ and $C=V\backslash (A\cup B)$. We  study  information flow between $A$ and $B$ when side information $C$ is considered. 
Mathematically, taking into account side information corresponds to conditioning on the side information.   As we outlined earlier   since we deal with the graph $V$ , we must use causal conditioning or we would break the symmetry between either $A$ or $B$ and $C$.
  This  leads to relate causal conditional directed information to the definitions \ref{GrangerDyn:def}, \ref{GrangerInst:def} and \ref{GrangerInstUn:def}. Since we have two possible definitions for instantaneous causality, we have two possible choices for using the side information as a conditioner: We may use the past $x_C^{k-1}=Dx_C^k$ or the past as well as the present $x_C^{k}$.

{\bf Conditioning on the past: }€  We evaluate $I(x_A^k \rightarrow x_B^k \big\| Dx_C^{k} )$. This can be written as 
\begin{eqnarray*}
I(x_A^k \rightarrow x_B^k\big\|  €  Dx_C^{k} ) = I(Dx_A^k \rightarrow x_B^k\big\|  €   Dx_C^k)+ I(x_A^k \rightarrow x_B^k\big\| Dx_A^{k} ,Dx_C^{k}  )
\end{eqnarray*}

We call the first term of this decomposition  $I(Dx_A^k \rightarrow x_B^k\big\|  €   Dx_C^k)$ the conditional transfer entropy between $A$ and $B$ given $C$. 
It is zero if and only if $I(x_A^{i-1} ; x_B(i) | € x_B^{i-1}, x_C^{i-1})=0$,   $\forall i$ or equivalently if and only if 
$P\big( x_B(k)  \big| x_A^{k-1},x_B^{k-1},x_C^{k-1} \big)  = P\big(x_B(k)  \big| x_B^{k-1},x_C^{k-1}  \big)$. In other words, according to definition \ref{GrangerDyn:def}, the conditional transfer entropy between $A$ and $B$ is zero if and only if
$A$ does not dynamically cause $B$. 

The second term $I(x_A^k \rightarrow x_B^k\big\| Dx_A^{k} ,Dx_C^{k}  )$ is zero if and only if $I(x_A(i) ; x_B(i) | € x_A^{i-1}, x_B^{i-1}, x_C^{i-1})=0$,   $\forall i$ and therefore, according to definition \ref{GrangerInst:def} if and only if $A$ does not unconditionally instantaneously cause $B$. We will refer to this measure as the unconditional instantaneous information exchange. Note that the ``unconditional'' term refers to the nature of the type of  independence the measure reveals.

{\bf Conditioning up to the present:} We evaluate $I(x_A^k \rightarrow x_B^k \big\| x_C^{k} )$. The idea is to find a decomposition in which both the conditional transfer entropy and a measure accounting for definition \ref{GrangerInstUn:def}
appears. Applying several times the chain rule for conditional mutual information, the defining term for the causal conditional directed information $I(x_A^k \rightarrow x_B^k \big\| x_C^{k} )$ verifies
\begin{eqnarray*}
I(x_A^i ; x_B(i) \big| x_B^{i-1}, x_C^{i})  &=&  I(x_C(i), x_A^i ; x_B(i) \big| x_B^{i-1}, x_C^{i-1} )  - I(x_C(i); x_B(i) \big| x_B^{i-1}, x_C^{i-1} ) \\
&=&   I( x_A^{i-1} ; x_B(i) \big| x_B^{i-1}, x_C^{i-1} ) +  I( x_C(i), x_A(i) ; x_B(i) \big| x_A^{i-1} , x_B^{i-1}, x_C^{i-1} )  \\ &-& I(x_C(i); x_B(i) \big| x_B^{i-1}, x_C^{i-1} ) \\
&=& I( x_A^{i-1} ; x_B(i) \big| x_B^{i-1}, x_C^{i-1} ) + I( x_A(i) ; x_B(i) \big| x_A^{i-1} , x_B^{i-1}, x_C^{i} )\\
&+& I( x_C(i) ; x_B(i) \big| x_A^{i-1} , x_B^{i-1}, x_C^{i-1} ) - I(x_C(i); x_B(i) \big| x_B^{i-1}, x_C^{i-1} ) 
\end{eqnarray*}
Summing over $i$ we get the conditional causal directed information 
\begin{eqnarray}
I(x_A^k \rightarrow x_B^k \big\| x_C^{k} ) = I(Dx_A^k \rightarrow x_B^k \big\| Dx_C^{k} ) +
I(x_A^k \rightarrow x_B^k \big\| Dx_A^k, x_C^{k} ) +\Delta I (Dx_C^k \rightarrow x_B^k \big\| Dx_A^k, Dx_C^{k} ) 
\label{decomp:eq}
\end{eqnarray}
The term $I(x_A^k \rightarrow x_B^k \big\| Dx_A^k, x_C^{k} )$ is called the conditional instantaneous information exchange. It is equal to zero if and only if definition \ref{GrangerInstUn:def} is verified, that is if and only if $A$ does not instantaneously cause $B$. We recover in the decomposition the conditional transfer entropy accounting for dynamical causality. The surprise   arises    from an extra-term  in eq. (\ref{decomp:eq})  defined as
\begin{eqnarray*}
\Delta I (Dx_C^k \rightarrow x_B^k \big\| Dx_A^k, Dx_C^{k} ) = I (Dx_C^k \rightarrow x_B^k \big\| Dx_A^k, Dx_C^{k} ) -
 I (Dx_C^k \rightarrow x_B^k \big\| Dx_C^{k} )
\end{eqnarray*}
This term is also measuring an instantaneous quantity.   It is the difference between  two different natures  of instantaneous coupling: The first  term  $I( x_C(i) ; x_B(i) \big| x_A^{i-1} , x_B^{i-1}, x_C^{i-1} )$   describes   intrinsic coupling in the sense it does not depend on other parties than $C$ and $B$; The second coupling  term  expressed by $ I(x_C(i); x_B(i) \big| x_B^{i-1}, x_C^{i-1} )$ is  relative to    extrinsic coupling since it measures the instantaneous coupling at time $i$ created by other variables than $B$ and $C$.

The conclusion is the following: causal directed information is the right measure to assess information flow in Granger causality graphs if the unconditional definition is adopted for instantaneous causality. In this case, causal directed information $I(x_A^k \rightarrow x_B^k \big\| Dx_C^{k} ) $ is zero if and only if there is no causality from $A$ to $B$. If not zero, we must evaluate the conditional transfer entropy and the unconditional instantaneous information exchange to assess dynamical and instantaneous causality. However, as shown by Eichler \cite{Eich07,Eich11}, the graphs obtained in this case do not have nice properties since the instantaneous graph is not a conditional dependence graph.

On the other hand, if we adopt definition \ref{GrangerInst:def} for instantaneous causality, we do not have the same nice decomposition, and $I(x_A^k \rightarrow x_B^k \big\| x_C^{k} ) $ cannot be used to check non causality. However, we have shown that the correct measures to assess dynamical and instantaneous causality are respectively the conditional transfer entropy $I(Dx_A^k \rightarrow x_B^k \big\| Dx_C^{k} ) $ and the instantaneous information exchange $I(x_A^k \rightarrow x_B^k \big\| Dx_A^k, x_C^{k} )$.

\section{Illustrations}

This section is devoted to the practical application of the previous results. We begin by discussing estimation issues and illustrate the key ideas on synthetic examples. 

\subsection{Estimation issues}

The estimator we use are based on Leonenko's $k$-nearest neighbour estimator of the entropy. Let $x_i, i=1,\ldots, N$
$N$ observations of some random vector $x$ taking values in $\R^n$. Then Leonenko's estimator for the entropy  reads 
\cite{GoriLMN05}
\begin{eqnarray*}
\widehat{H}_k (x) =\frac{1}{N} \sum_{i=1}^{N} \log\left(  (N-1) C_{k} V_n d\big(x_i,x_{i(k)}\big)^n \right) 
\end{eqnarray*}
In this expression, $d:\R^n\times \R^n \longrightarrow \R^+$ is a metric. $x_{i(k)}$ is defined to be the $k$th nearest neighbor of $x_i$. $V_n$ is the volume of the unit ball for the metric $d$; $C_k=\exp(-\psi(k))$, $\psi(.)$ is the digamma function defined as the derivative of the logarithm of the Gamma function. It is shown in \cite{GoriLMN05} that this estimator converges in the mean square sense to the entropy of the random vector $x$ (under the i.i.d. assumption of the $x_i$) for any values of $k$ lower than $N-1$.

{\bf Estimation of the conditional mutual information.}
To estimate a directed information, we need to estimate conditional mutual information $I(a,b|c)=H(b,c)+H(a,c)-H(a,b,c)-H(c)$.
Thus estimating the conditional mutual information can be done using four  applications  of Leonenko's estimator. Although it is  asymptotically unbiased, Leonenko's estimator is biased  for   finite sample size, and the bias depends on the dimension of the underlying space. Apart from the fact that a plug-in estimator would suffer from a high variance, the bias of the entropy estimators will therefore not cancel out. 


 A  smart  idea to circumvent this problem was proposed by Kraskov in 2004 for the mutual information case, and later extended by Frenzel and Pompe for the conditional mutual information case  \cite{KrasSG04,FrenP07}. The idea relies on two facts: The estimator is valid for any metric, and the estimator converges for any $k\leq N-1$. The idea is then to use as a metric in the product space the maximum of the metric used on the marginal spaces. This determine as a scale in the product space the distance $d\big(x_i,x_{i(k)}\big)$ between $x_i$ and its $k$th nearest neighbor. This distance is then used on the marginals to determine  $k'$ for which $d\big(x_i,x_{i(k)}\big)$ is the distance between $x_i$ projected on the marginal to its $k'$ nearest neighbour. 

{\bf Estimation of the directed information.}
Estimation requires that the processes studied are ergodic and stationary. Without these basic assumptions, nothing can really be done. The goal is to estimate the transfer entropy and the instantaneous information exchange. When dealing with monovariate signals $x_A(k)=x(k)$ and $x_B(k)=y(k)$, and with side information $x_C(k)$ the information measures read
\begin{eqnarray*}
I(Dx^k \rightarrow y^k \big\| Dx_C^k) &= &\sum_{i=1}^k I(x^{i-1}; y(i) \big| y^{i-1}, x_C^{i-1}) \\
I(Dx^k \rightarrow y^k \big\| Dx^k,x_C^k) &= &\sum_i I(x(i); y(i) \big| y^{i-1}, x^{i-1}, x_C^{i}) \\
\end{eqnarray*}

For stationary sequences, it is convenient to consider the rate of growth of these measures. Indeed, the measures are often linearly increasing with $k$. Thus the rate is defined as the asymptotic linear growth rate. Furthermore, following the proof in \cite{Kram98} for the directed information (or in \cite{CoveT93} €   for the entropy), it can be shown that 
\begin{eqnarray*}
\lim_{k\rightarrow+\infty} \frac{1}{k} I(Dx^k \rightarrow y^k \big\| Dx_C^k) &= &\lim_{k\rightarrow+\infty}I(x^{k-1}; y(k) \big| y^{k-1}, x_C^{k-1}) \\
\lim_{k\rightarrow+\infty} \frac{1}{k} I(Dx^k \rightarrow y^k \big\| Dx^k,x_C^k) &= &  \lim_{k\rightarrow+\infty} I(x(k); y(k) \big| y^{k-1}, x^{k-1}, x_C^{k}) 
\end{eqnarray*}

Suppose now that we are dealing we finite order joint Markov sequences. Then by working with vectors, we can represent signal using an order 1 Markov multivariate process. We thus assume that $(x,y,x_C)$ is a Markov process of order 1. Under this assumption and stationarity, we have 
\begin{eqnarray*}
\lim_{k\rightarrow+\infty}I(x^{k-1}; y(k) \big| y^{k-1}, x_C^{k-1}) &=& I(x(1); y(2) \big| y(1), x_C(1))   \\
  \lim_{k\rightarrow+\infty} I(x(k); y(k) \big| y^{k-1}, x^{k-1}, x_C^{k}) &=&I(x(2); y(2) \big| y(1), x(1), x_C^{2}) 
\end{eqnarray*}
and in this case, we can estimate the conditional transfer entropy and the instantaneous information exchange
from data.

Practically, from two times series $x$ and $y$ and a pool of others $x_C$, we create from the signals the realizations of the vectors
$x(1)_i=x_{i-d}^{i-1}$, $y(1)_i=y_{i-d}^{i-1}$, $x_{C}(1)_i=x_{C,i-d}^{i-1}$ and ${x_{C}^2}_i=x_{C,i-d}^{i}$, and estimate 
$I(x(1); y(2) \big| y(1), x_C(1)) $ and $I(x(2); y(2) \big| y(1), x(1), x_C^{2}) $ using these realizations and the $k$-nn estimators described above \cite{KrasSG04,FrenP07}. This approach has already been described in \cite{ViceWLP11} for the transfer entropy. 

\subsection{Synthetic examples}

We develop here two synthetic examples to illustrate the key ideas developed in the paper. In the first example, we stress the importance of causal conditioning using a simple causality chain. The second example is a particular instance of the example developed in the second section of the paper, for which we estimate dynamical and instantaneous causality measures.

\subsection{A chain}

\begin{figure}[p]
\begin{center}
 \includegraphics[scale=.8]{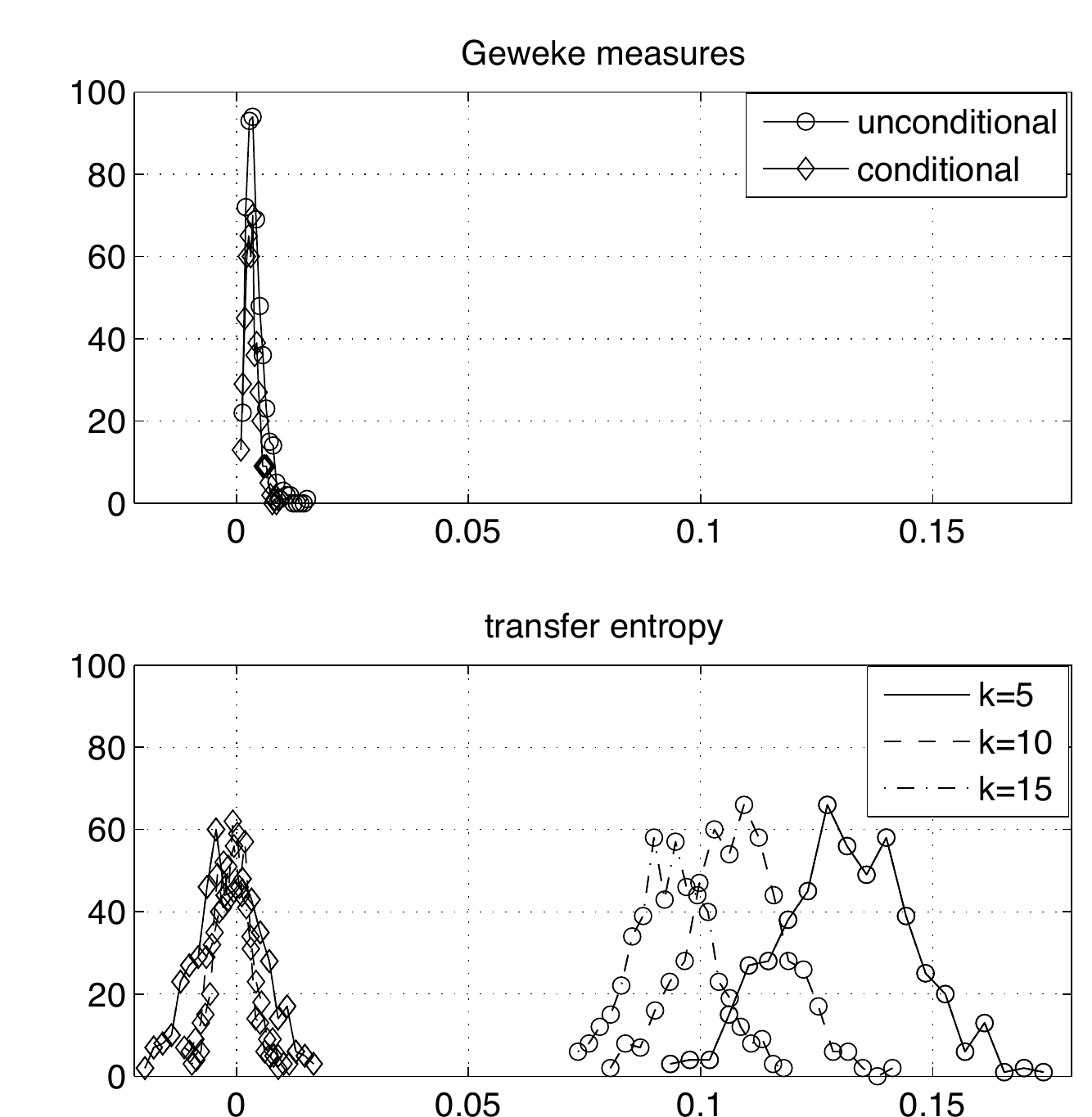}
 \end{center}
 \caption{Dynamical causality analysis from $x$ to $z$ in the first example. Top: linear analysis using Geweke's measures.
 Both  conditional and unconditional measures lead  to conclude   that $x \not \rightarrow z$. Bottom: directed information theoretic analysis. The three different types of histograms correspond to three different choice of the number of nearest neighbours $k$ for the estimation. As can be seen, the variance decreases with $k$ but the bias increases. From the  transfer entropy, since $I(x\rightarrow z)>0$, we obtain $x\rightarrow z$ whereas the condititional transfer entropy leads to $x\rightarrow z$ since $I(x\rightarrow z\| y)=0$.}
 \label{chain:fig}
\end{figure}

Consider the following three dimensional example, in which the noises are i.i.d. and independent of each other.
\begin{eqnarray*}
 x_t & =&b x_{t-1} + \varepsilon_{x,t}  \\
 y_t &= &c y_{t-1}  +d_{xy}x_{t-1}^2+ \varepsilon_{y,t} \\
 z_t &=& d z_{t-1} +c_{yz} y_{t-1} + \varepsilon_{z,t} 
\end{eqnarray*}
where $a=0.2,b=0.5,c=0.8, d_{xy}=0.8,c_{yz}=0.7$.
Firstly, we evaluate Geweke's measure based on linear prediction error \cite{Gewe82,Gewe84} (logarithm of the {\it ratio} 
between variances of linear prediction). The measures are evaluated on 100 independent realizations of length 3000 samples of the processes. They are depicted in figure (\ref{chain:fig})
in the form of histograms. 
As can be seen, the histogram for the conditional Geweke measure $F_{xz\|y}$ has the same support as the histogram of the unconditional measure $F_{xz}$. Therefore, we have an example where linear Granger causality gives the same answer whether conditional or not:  $x$ does not dynamically cause $z$ (conditional or not to $y$).

We then evaluate the transfer entropy $I(Dx\rightarrow z)=I(x_{t-2}^{t-1}; z_t \big| z_{t-2}^{t-1} )$ and the conditional transfer entropy  $I(Dx\rightarrow z \big\| Dy )=I(x_{t-2}^{t-1}; z_t \big| z_{t-2}^{t-1} ,y_{t-2}^{t-1})$ on the same data sets. The results are depicted in the bottom of figure (\ref{chain:fig}). We see that the histograms of the conditional measure is clearly centered around 0 whereas the histogram for the unconditional measure has clearly a non overlapping support. Therefore  we conclude  that when side information is not taken into account, $x$ causes $z$, whereas including $y$ as side information reverses the conclusion. Therefore, the existing link from $x$ to $z$ passes through $y$. In the plot of the transfer entropy, we present the histograms of the measures for three different values of $k$, the number of nearest neighbors considered by the estimation. As seen and reported  in   \cite{AmblZMC08}, there is a trade-of between bias and variance as a function of $k$.  The  present   lack of precise theoretical analysis does not allow  to optimize this trade-off in order to choose $k$ (see however \cite{SricH11} for a work going in this direction). However, numerical simulations have shown that $k$ should be chosen small as the dimension of the space increases.

\subsection{A four dimensional complete toy}

We come back to the example described in section \ref{detailedexample:ssec}.  Below we  provide an explicit form to the functional links
\begin{eqnarray}
\left\{
 \begin{array}{ccl}
 w_t &=& a w_{t-1}   +\alpha z_{t-1} +    e x_{t-1}^2  +\varepsilon_{w,t} \\
 x_t & =&b x_{t-1} +                               f z^2_{t-1}+   \varepsilon_{x,t} \\
 y_t &=& c y_{t-1} + \beta x_{t-1}+         g x_{t-1}^2+ \varepsilon_{y,t} \\
 z_t &=& d z_{t-1} +\gamma  w_{t-1} +                       \varepsilon_{z,t} \\
 \end{array}
 \right. 
 \label{exemple4d:eq}
\end{eqnarray}
and we recall that the noise sequence is white with covariance given by (\ref{covariance:eq}). For the purpose of the example, we  set  $\rho_1=0.66, \rho_2=0.55$ and $\rho_3=0.48$. To mimic a real experiment we have simulated a long time series from which $N_b=100$ consecutive blocks of 3000 samples each was used to generate the realizations of the process. Thus,
all the information measures needed were evaluated on these blocks. Furthermore, we perform random permutations to simulate the independence situation called $H_0$. Precisely, when estimating $I(a;b|c)$ from samples $a_i,b_i,c_i$, the permutation is done on the $b_i$'s. Indeed if  $b$ is   independent from  $a$ and $c$  then  $I(a;b|c)=0$. For example, when estimating  the   transfer entropy $I(x_{t-2}^{t-1}; z_t \big| z_{t-2}^{t-1} )$ we use permutation for $z_t$ but not for $z_{t-2}^{t-1}$. Thus for each block,  two measures are actually performedcorresponding to the one that needs to be evaluated and another one for which  $H_0$ hypothesis is forced . The $N_b$  results  under $H_0$ allow to  evaluate the threshold $\eta_{ij}$ over which only $\alpha \%$ of false positive decisions (there is a link from $i$ to $j$) will be taken. Practically we  set  $\alpha=10\%$.
 Since for this toy problem 12 dependence pairwise tests need to be made,  the Bonferronni\footnote{
{\it i.e.} $\alpha$ is replaced with $\alpha/12$ to ensure a family false positive rate less than $\alpha \%$. Note that Bonferronni correction is known to be very conservative, and less conservative procedure such as False Discovery Rate control could be easily adopted.}
correction is applied to the threshold, in order to maintain the family-wise global false detection rate.   9 different measures were tested on this example: 
\begin{enumerate}
\item Geweke's instantaneous causality measure 
$$
F_{x y}=\lim_{n\rightarrow +\infty} {\varepsilon(x_n | x^{n-1},y^{n-1})\over \varepsilon(x_n | x^{n-1},y^{n})}.
$$
where $\varepsilon(x  | z )$ is the variance of the error in the linear estimation of $x$ from $y$.
\item Geweke's conditional instantaneous causality measure 
$$
F_{x y}=\lim_{n\rightarrow +\infty} {\varepsilon(x_n | x^{n-1},y^{n-1}, (w,z)^n)\over \varepsilon(x_n | x^{n-1},y^{n},(w,z)^n)}.
$$
\item Geweke's dynamical  causality measure 
$$
F_{x\rightarrow y}=\lim_{n\rightarrow +\infty}  {\varepsilon(x_n | x^{n-1})\over \varepsilon(x_n | x^{n-1},y^{n-1})}.
$$
\item Geweke's conditional dynamical  causality measure 
$$
F_{x\rightarrow y}=\lim_{n\rightarrow +\infty} {\varepsilon(x_n | x^{n-1}, (w,z)^{n-1})\over \varepsilon(x_n | x^{n-1},y^{n-1},(w,z)^{n-1})}.
$$
\item Instantaneous information exchange $I(x^n \rightarrow y^n \| Dx^n)$.

\item Instantaneous unconditional information exchange $I(x^n \rightarrow y^n \| Dx^n, Dw^n,Dz^n)$.

\item Instantaneous conditional information exchange $I(x^n \rightarrow y^n \| Dx^n, w^n,z^n)$.

\item Transfer entropy $I(Dx^n  \rightarrow y^n )$.
\item Conditional transfer entropy $I(Dx^n  \rightarrow y^n \| Dw^n,Dz^n)$.
\end{enumerate}

\begin{figure}[p]
\begin{center}
 \includegraphics[scale=.75]{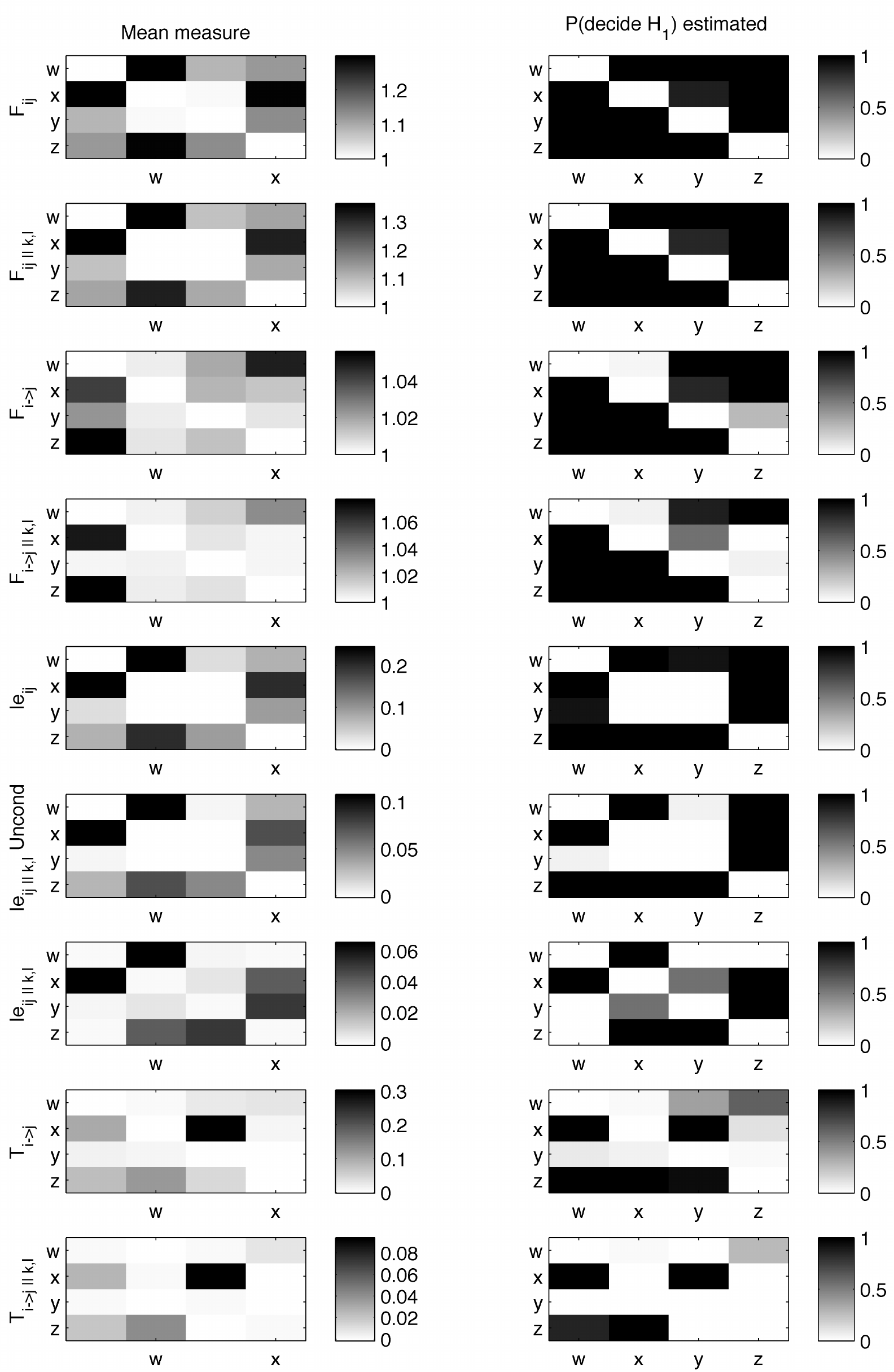}
 \end{center}
 \caption{Measures calculated from example 2. From top to bottom, instantaneous causality and conditional instantaneous causality  Geweke's measures, dynamical causality  and conditional dynamical causality Geweke's measure, instantaneous information exchange, unconditional instantaneous  information exchange, conditional instantaneous  information exchange, transfer entropy and finally conditional transfer entropy. The left column is the mean measure calculated over 100 realizations of 3000 samples each. The right column represented the number of time the corresponding measure exceeds a threshold chosen to ensure a family false positive probability of 10 \% (using Bonferronni correction).  }
 \label{4Dexample:fig}
\end{figure}

Geweke's measures are based on linear estimation. Note that they  take values larger   than one when $x$ causes $y$ and equal to one otherwise. We  stressed  that (up to a log) Geweke's measure are the Gaussian version of directed information measures discussed here \cite{AmblM09,AmblM11:it}.  Information measures were  estimated   using the  appropriate conditional mutual information definitions,  with time lag windows of length 2, {\it e.g.} € the conditional transfer entropy $I(Dx^n  \rightarrow y^n \| Dw^n,Dz^n)$ is approximated by the estimation of
 $I(x_{t-2}^{t-1}; y_t \big| y_{t-2}^{t-1} ,w_{t-2}^{t-1},z_{t-2}^{t-1})$. The results are depicted in figure (\ref{4Dexample:fig}). The  measures  1 to 9 are depicted  from top to  bottom. The left column  represents  the matrix of the measures averaged over the $N_b$ blocks. The right column  represents  the matrix of estimated probabilities of deciding that there is a link between two nodes.
 To estimate that there is a link, we use the threshold $\eta_{ij}$ discussed above. To evaluate Geweke's measure we perform a linear prediction using 10 samples in the past and evaluate the variance of the errors.
 Note that the diagonal of all these matrices is put to arbitrarily to zero since the diagonal is not informative in this study.

The main conclusions to be drawn from this  experiments  are the following.
\begin{itemize}
\item The linear analysis, whether causally conditional or not, implemented using Geweke's measures, fails to retrieve the structure of the causality graphs. 

\item The instantaneous information exchange must be causally conditioned, since  the results given in the fifth line of the figure ( $Ie_{ij}$ ) does not reveal the exact nature of the dependencies.

\item The importance of the horizon of causal conditioning appear in the 6th and 7th line where we plot the results for respectively the unconditional and conditional instantaneous information exchange. The measures are correctly estimated, since the probability of assigning links is very high as shown in the right column: the form of the matrices are correct. We recover the form of the covariance matrix of the noise using the unconditional form whereas we recover the form of the precision matrix using
the conditional form of the instantaneous information exchange. Note on this example a rather low probability of estimating the link between  $x$ and $y$ in the conditional form.

\item The causality graph needs causal conditioning to be correctly inferred, as revealed by the two last measures. 
However note again a rather low probability of estimating correctly the link from $w$ to $z$, a difficulty  clearly due to the low coupling constant $\gamma$ existing in this direction. Trying to increase this coupling to study the sensitivity is unfortunately impossible since increasing 
slightly $\gamma$ destabilize the system. 

\end{itemize}

\section{Conclusion}

In this paper, we have revisited and highlighted the links 
between directed information theory and Granger causality graphs. In the bivariate case, the directed information decomposes into the sum of  two contributions: the transfer entropy and the instantaneous information exchange. 
Each term  in this decomposition reveals a type of causality. Transfer entropy between  two processes (say $X$ and $Y$)  is zero if and only if there is no dynamical Granger  causality: the knowledge of the past of $X$ dos not lead to any improvement in the prediction quality of $Y$. Instantaneous information exchange quantifies the instantaneous link that may exist between  the two signals. 


In the multivariate case however,    instantaneous   causality gives rise to increased difficulties when relating  directed information theory to  the measures introduced in the bivariate case.  We have recalled that  two definitions of instantaneous causality may be given, depending on the time horizon selected in the consideration of side information. If the past of the side information is considered, instantaneous causality leads to a concept of independence graph models, whereas consideration of the present of the side information as well leads to a conditional graphical model. Preferring one of these definitions leads to a  rather a difficult choice, discussed in this paper:  Conditional graphs enjoy  nice Markov properties whereas unconditional graphs represent a preferred solution   in neuroscience, as it provides a better matches to the concept of functional connectivity \cite{Sporn10}.

 We have also shown that  if   independence graphs are considered, directed information causally conditioned to the past of the side information decomposes into the sum of the causally conditioned transfer entropy and the causally conditioned (independent) information exchange, directly extending the bivariate result.  This   decomposition however   does not longer hold  in  the other case. For the conditional graph  an extra term appears in the decomposition.  It further explains how instantaneous exchange   takes place    between the two signals of interest and the side information.

All this theoretical framework   finds some practical developments as illustrated on two synthetic examples. The estimators we used in this paper rely  on nearest neighbors   based entropy estimators. These estimators can be efficiently used  as long as  the dimensionality of the problems at hand is not high.

\section*{Acknowledgements} € P.O.A. is supported by a Marie Curie International Outgoing Fellowship from the European Community.

\bibliographystyle{elsarticle-num}








\end{document}